\documentstyle[12pt]{article}
\newfont{\ffont}{msym10}                        %%
\newcommand{\beq}{\begin{equation}}             %%
\newcommand{\eeq}{\end{equation}}               %%
\newcommand{\bqry}{\begin{eqnarray}}            %%
\newcommand{\eqry}{\end{eqnarray}}              %%
\newcommand{\bqryn}{\begin{eqnarray*}}          %%
\newcommand{\eqryn}{\end{eqnarray*}}            %%
\newcommand{\preprint}[1]{\begin{table}[t]      %%
            \begin{flushright}                  %%
            \begin{large}{#1}\end{large}        %%
            \end{flushright}                    %%
            \end{table}}                        %%
\newcommand{\PD}[2]                             %%
    {\frac{\partial^{#2}}{\partial #1^{#2}}}    %%
\begin{document}
\preprint{LA-UR-96-2715 \\ IASSNS-HEP-96/XX}
\title{On the Thermodynamics of \\ Hot Hadronic Matter}
\author{\\ L. Burakovsky\thanks{Bitnet: BURAKOV@QCD.LANL.GOV} \
\\  \\  Theoretical Division, T-8 \\  Los Alamos National  
Laboratory \\ Los
Alamos NM 87545, USA \\  \\  and  \\  \\
L.P. Horwitz\thanks{Bitnet: HORWITZ@SNS.IAS.EDU. On sabbatical leave from
School of Physics and Astronomy, Tel Aviv University, Ramat Aviv, Israel.
Also at Department of Physics, Bar-Ilan University, Ramat-Gan,  
Israel  } \
\\  \\ School of Natural Sciences \\ Institute for Advanced Study  
\\ Princeton
NJ 08540, USA \\}
\date{  }
\maketitle
\begin{abstract}
The equation of state of hot hadronic matter is obtained, by taking into 
account the contribution of the massive states with the help of the  
resonance
spectrum $\tau (m)\sim m^3$ justified by the authors in previous  
papers. This
equation of state is in agreement with that provided by the  
low-temperature
expansion for the pion intracting gas. It is shown that in this  
picture the
deconfinement phase transition is absent, in agreement with lattice gauge
calculations which show the only phase transition of chiral symmetry 
restoration. The latter is modelled with the help of the  
restriction of the
number of the effective degrees of freedom in the hadron phase to that
of the microscopic degrees of freedom in the quark-gluon phase,  
through the
corresponding truncation of the hadronic resonance spectrum, and  
the decrease
of the effective hadron masses with temperature, predicted by Brown  
and Rho.
The results are in agreement with lattice gauge data and show a smooth
crossover in the thermodynamic variables in a temperature range  
$\sim 50$ MeV.
\end{abstract}

\bigskip
\section{Introduction}
One of the main goals of experiments with high-energy nuclear collisions
is to produce and to study hadronic matter, in particular, trying to
reach conditions at which the phase transitions into the quark-gluon
plasma phase can take place [1--5]. The physics of hot hadronic matter
has not been studied much, although such matter is already produced in
current experiments in high-energy physics.

The experimental data on multiple hadron production obtained in recent
years are in agreement with the main consequences of the theory
formulated by Landau \cite{Lan} over 40 years ago. However, with a
sufficiently quantitative approach, it becomes necessary to consider a
number of physical effects which bring about certain modifications of the
results obtained in the fundamental work [6]. For instance, the solution
of the equations of motion obtained by Landau differs from a numerical
calculation of Milekhin \cite{Mel1}, as a result of an inaccurate
estimate. A more accurate analytic solution was given in \cite{Shu1}.

The equation of state in Landau's work is taken to be $p=\rho /3 $ (where
$p$ is the pressure and $\rho $ is the energy density), corresponding to
an ultra-relativistic gas. However, the CERN and FNAL colliders provide 
proton-antiproton collisions with the total center-of-mass energy of the 
order of 1 TeV, corresponding to initial temperatures  
$T\stackrel{<}{\sim }
10$ GeV for which the interaction of the hadrons is strong and has  
mainly a
resonant character, the masses of the resonances being comparable  
with the
temperature. Thus, hadronic matter under these conditions is neither
an ideal nor an ultrarelativistic gas.

Corrections in the equation of state due to the interaction of the
hadrons have been discussed in the literature for the last three
decades [1,5,8-16]. These considerations were based mostly on a
phenomenological model in which the Landau theory is applied to all
particles except the leading ones, i.e., the fragments of the initial
particles, whose characteristics are taken directly from experimental
data. The framework for the latter considerations is the QCD phase
transition of hadronic matter into the quark-gluon plasma.

Hot hadronic matter, in which volumes per particle are a few cubic
fermis, is certainly made out of individual hadrons. It is clear that,
at low temperatures $T<<m_\pi ,$ one has a very rare (and therefore
ideal) gas of the lightest hadrons, the pions. As the temperature is
raised and the gas becomes more dense, one should take into account
interactions among the particles. This was done using the following three
approaches: (i) the low-temperature expansion, (ii) the resonance gas,
(iii) the quasi-particle gas.

The first approach for the pion gas is based on the Weinberg theory of
pion interactions \cite{Wei}, which uses the non-linear Lagrangian
containing all processes quadratic in pion momentum. Its first
application to calculation of the thermodynamic parameters of the pion
gas was done in ref. [1]. One of the important consequences of this work 
was that, after isospin averaging, all corrections quadratic in momenta
cancel each other, and corrections proportional to the pion mass
(Weinberg $\pi \pi $ scattering lengths) are nearly compensated,
producing negligible corrections at the 1\% level. Corrections of the
second order in the Weinberg Lagrangian were also estimated in [1], and
a much more systematic study of the problem including quartic terms in
the mesonic Lagrangian were made in ref. \cite{GL}. It has been  
shown that
taking into account the pion rescattering leads to corrections  
quartic in
momenta which, after isospin averaging, give a nonzero correction to the
pressure, providing a scale for the temperatures at which  
deviations from the
ideal pion gas formula become noticeable:
\beq
p=\frac{\pi  
^2}{30}T^4\left[\;1+\left(\frac{T}{T_{int}}\right)^4\;\right],
\;\;\;T_{int}\simeq 150\;{\rm MeV}.
\eeq
Without going into discussion of these works we remark that the  
applicability
of this approach is limited by temperatures $T<100$ MeV, for which  
typical
collision energies are significantly below resonance. However, such  
$T$ are
lower than even the lowest temperature available in experiments,  
because the
so-called break-up temperatures are typically $T\simeq 120-150$ MeV.

The idea of resonance gas was first suggested by Belenky and Landau
\cite{BL}, as early as 1956. They used the Beth-Uhlenbeck method
\cite{BU}, well known in the theory of the non-ideal gases \cite{LL},
the main idea of which consists in the calculation of the number of  
states
within the normalization volumewith the scattering phase shifts  
taken into
account. At large distances $r$ between the particles, their wave  
function is
given by $\psi _\ell (r)\approx \sin (pr+\delta _\ell )/r,$ and the  
boundary
condition $\psi _\ell (R)=0$ at the boundary of the normalization  
volume picks
out the states with momenta $pR+\delta _\ell (p)=\pi n.$ The  
replacement of
the summation over $n$ in the statistical sum by an integral over  
$p$ gives
rise to the following expression ($Z$ is only a part of the  
statistical sum
connected with the relative motion of the particles):
\beq
Z=\frac{1}{\pi }\sum _\ell (2\ell +1)\int _0^\infty dp\;\left(R+\frac{
d\delta _\ell (p)}{dp}\right)e^{-E(p)/T}.
\eeq
If there is a resonance in the scattering at a certain momentum  
$p_0,$ then
$\delta _\ell (p)$ changes rapidly in the neighborhood of $p_0.$ It  
is easily
seen that the contribution of the resonance to the integral (2) is  
the term
$\exp \;(-E(p_0)/T),$ the same as for a bound state with that energy. 
Therefore, resonances and stable particles should be taken into  
account on an
equal footing in the thermodynamic characteristics of hot hadronic  
matter;
e.g., the formulas for the pressure and energy density in a  
resonance gas read
(we neglect the chemical potential for simplicity)
\beq
p=\sum _ip_i=\sum _ig_i\frac{m_i^2T^2}{2\pi ^2}\sum _{r=1}^\infty  
(\pm 1)^{r+1}
\frac{K_2(rm_i/T)}{r^2},
\eeq
\beq
\rho =\sum _i\rho _i,\;\;\;\rho _i=T\frac{dp_i}{dT}-p_i,
\eeq
where $+1$ $(-1)$ corresponds to the Bose-Einstein (Fermi-Dirac)  
statistics,
and $g_i$ are the corresponding degeneracies ($J$ and $I$ are spin and
isospin, respectively), $$g_i=\left[
\begin{array}{ll}
(2J_i+1)(2I_i+1) & {\rm for\;non-strange\;mesons} \\
4(2J_i+1) & {\rm for\;strange}\;(K)\;{\rm mesons} \\
2(2J_i+1)(2I_i+1) & {\rm for\;baryons}
\end{array} \right. $$
These expressions may be rewritten with the help of a {\it  
resonance spectrum,}
\beq
p=\int _{m_1}^{m_2}dm\;\tau (m)p(m),\;\;\;p(m)\equiv  
\frac{m^2T^2}{2\pi ^2}
\sum _r(\pm 1)^{r+1}\frac{K_2(rm/T)}{r^2},
\eeq
\beq
\rho =\int _{m_1}^{m_2}dm\;\tau (m)\rho (m),\;\;\;\rho (m)\equiv
T\frac{dp(m)}{dT}-p(m),
\eeq
normalized as
\beq
\int _{m_1}^{m_2}dm\;\tau (m)=\sum _ig_i,
\eeq
where $m_1$ and $m_2$ are the masses of the lightest and heaviest  
species,
respectively, entering the formulas (3),(4).

In both the statistical bootstrap model \cite{Hag,Fra} and the dual  
resonance
model \cite{FV}, a resonance spectrum takes on the form
\beq
\tau (m)\sim m^a\;e^{m/T_0},
\eeq
where $a$ and $T_0$ are constants. The treatment of hadronic  
resonance gas by
means of the spectrum (8) leads to a singularity in the thermodynamic 
functions at $T=T_0$ \cite{Hag,Fra} and, in particular, to an  
infinite number
of the effective degrees of freedom in the hadron phase, thus hindering a
transition to the quark-gluon phase. Moreover, as shown by Fowler  
and Weiner
\cite{FW}, an exponential mass spectrum of the form (8) is  
incompatible with
the existence of the quark-gluon phase: in order that a phase  
transition from
the hadron phase to the quark-gluon phase be possible, the hadronic  
spectrum
cannot grow with $m$ faster than a power.

In 1972, Shuryak \cite{Shu1}, in place of (8), used the simple power
parametrization
\beq
\tau (m)\sim m^k,
\eeq
which ``describes experiment in the mass region $0.2-1.5$ GeV for  
$k\approx
3.$'' Since $\tau (m)$ is a rapidly growing function, the main  
contribution to
integrals of type (10) is given by the mass region in which $\exp \;(m/T)
>>1,$ and the difference in properties of bosons and fermions is
irrelevant. Upon substitution of (9) into (5) and (6) one finds that the
pressure and energy density are proportional to $T^{k+5}.$  
Therefore, the
velocity of sound $c_s,$ defined by
\beq
c^2_s=\frac{dp}{d\rho },
\eeq
turns out to be a temperature independent constant, and the equation of
state belongs to the class considered by Milekhin \cite{Mel2}. For this
case, the expressions for energy and pressure are as follows,
\beq
\rho =\lambda T^{k+5},\;\;\;p=\frac{\lambda  
}{k+4}T^{k+5},\;\;\;c_s^2=\frac{1
}{k+4},
\eeq
where $\lambda $ is a certain constant. Consequently, for $k\approx 3$
\cite{Shu1}, $c_s^2\approx 0.14.$

In 1975, Zhirov and Shuryak \cite{ZS} have calculated the velocity  
of sound,
Eq. (10), directly from Eqs. (5),(6), and found that at first it  
increases
with $T$ very quickly and then saturates at the value of  
$c_s^2\simeq 1/3$ if
only the pions are taken into account, and $c_s^2\simeq 1/5$ if  
resonances up
to $M\sim 1.7$ GeV are included, hence favoring $k\approx 1$ in  
(11) and the
equation of state $p,\rho \sim T^6,$ $p=\rho /5.$ Such an equation  
of state
was suggested by Shuryak in his book \cite{QCD} of 1988 (and  
referred to also
in [1]), by fitting the real resonance spectrum,
\beq
p\simeq \left(20\;{\rm GeV}^{-2}\right)T^6,
\eeq
and called the ``realistic'' equation of state of hot hadronic  
matter. It
gives $c_s^2=0.20,$ in agreement with the theoretical models considered  
in earlier works \cite{SW,Gor,And}.

In ref. \cite{Shu2} Shuryak, in order to describe the behavior of the
energy density with temperature $\rho \propto T^6,$ proposed to consider
hot hadronic matter as a gas of quasiparticles, which have quantum
numbers of the original mesons, but with dispersion relations
modified by the interaction with matter (similarly to Landau's idea
of ``rotons'', to explain the growth of the energy density of liquid
$^4He$ with temperature more rapidly than $T^4).$
Since in such an approach hadrons are included as physical
degrees of freedom of hot hadronic matter, they are assumed not to be
absorbed too strongly, i.e., to be good quasiparticles. In ref.
\cite{Shu2} an attempt was made to guess what these dispersion relations
should be, in order to obtain the expected behavior of the energy density
with temperature. In \cite{Shu4} these dispersion relations were
calculated in explicit form.

It may seem strange that the different approaches (i.e., (i) and  
(ii) out of
the three mentioned above, the approach (iii) was invented in order  
to justify
the results given by (ii)) should lead to different results, viz.,  
Eqs. (1)
and (12); moreover, it is not clear which result, if it does,  
corresponds to
the genuine thermodynamics of the hadronic resonance gas. In this  
article we
shall show that with the correct form of the hadronic resonance  
spectrum which
has been established by the authors in a series of papers
\cite{spectrum,su4,linear}, the equation of state of hot hadronic  
matter takes
on the form (1), but reduces to (12) in the case of only lower mass  
resonances
being taken into account.

\section{Hadronic resonance spectrum and the equation of state}
In our previous work \cite{spectrum} we considered a model for a  
transition
from a phase of strongly interacting hadron constituents, described by a 
manifestly covariant relativistic statistical mechanics which  
turned out to be
a reliable framework in the description of realistic physical systems 
\cite{mancov}, to the hadron phase described by a resonance  
spectrum, Eqs.
(5),(6). An example of such a transition may be a relativistic high 
temperature Bose-Einstein condensation studied by the authors in ref. 
\cite{cond}, which corresponds, in the way suggested by Haber and Weldon 
\cite{HW}, to spontaneous flavor symmetry breakdown, $SU(3)_F\rightarrow 
SU(2)_I\times U(1)_Y,$ upon which hadronic multiplets are formed,  
with the
masses obeying the Gell-Mann--Okubo formulas \cite{GMO}
\beq
m^\ell =a+bY+c\left[ \frac{Y^2}{4}-I(I+1)\right];
\eeq
here $I$ and $Y$ are the isospin and hypercharge, respectively,  
$\ell $ is 2
for mesons and 1 bor baryons, and $a,b,c$ are independent of $I$  
and $Y$ but,
in general, depend on $(p,q),$ where $(p,q)$ is any irreducible  
representation
of $SU(3).$ Then the only assumption on the overall degeneracy  
being conserved
during the transition leads to the unique form of a resonance  
spectrum in the
hadron phase:
\beq
\tau (m)=Cm,\;\;\;C={\rm const},
\eeq
in agreement with the results of Zhirov and Shuryak \cite{ZS} found on a 
phenomenological ground. We have checked the coincidence of the  
results given
by a linear spectrum (14) with those obtained directly from Eq. (3)  
for the
actual hadronic species with the corresponding degeneracies, for all 
well-established hadronic multiplets, both mesonic and baryonic,  
and found it
excellent \cite{spectrum}. Therefore, the fact established  
theoretically that
a linear spectrum is the actual spectrum in the description of  
individual
hadronic multiplets, finds its experimental confirmation as well. In our 
recent papers \cite{su4,linear} we have shown that a linear  
spectrum of an
individual meson nonet is consistent with the Gell-Mann--Okubo mass  
formula
\beq
m_1^2+3m_8^2=4m_{1/2}^2
\eeq
(in fact, this formula may be derived with the help of a linear spectrum 
\cite{su4}), and leads to an extra relation for the masses of the  
isoscalar
states, $m_{0^{'}}$ and $m_{0^{''}},$ (of which $0^{'}$ belongs to  
a mostly
octet),
\beq
m_{0^{'}}^2+m_{0^{''}}^2=m_0^2+m_8^2=2m_{1/2}^2,
\eeq
with $m_1,m_{1/2},m_8,m_0$ being the masses of the isovector,  
isospinor, and
isoscalar octet and singlet states, respectively, which for an  
almost ideally
mixed nonet reduces to \cite{su4,linear}
\beq
m_{0^{''}}^2\simeq m_1^2,\;\;\;m_{0^{'}}^2\simeq 2m_{1/2}^2-m_1^2.
\eeq
We have checked the relation (17) in ref. \cite{linear} and shown  
to hold with
an accuracy of up to $\sim $3\% for all well-established nonets. In ref. 
\cite{su4} we have generalized a linear spectrum to the case of  
four quark
flavors and derived the corresponding Gell-Mann--Okubo mass formula  
for an
$SU(4)$ meson hexadecuplet, in a good agreement with the experimentally 
established masses of the charmed mesons. In ref. \cite{enigmas} we have 
applied a linear spectrum to the problem of establishing the correct 
$q\bar{q}$ assignment for the problematic meson nonets, like the scalar, 
axial-vector and tensor ones, and separating out non-$q\bar{q}$  
mesons. In
this paper we shall apply a resonance spectrum to the derivation of the 
thermodynamic characteristics of hot hadronic matter.

Let us first consider the case of the meson resonances alone. For  
this case,
the normalization constant $C$ of a linear spectrum (14) was  
established in
ref. \cite{spectrum}: for a nonet, one has 9 isospin degrees of  
freedom lying
in the interval $(m_{0^{''}}\simeq m_1,\;m_{0^{'}}).$ Therefore, Eq. (7) 
gives $$C\int _{m_{0^{''}}}^{m_{0^{'}}}dm\;m=9,$$ and hence
\beq
C=\frac{18}{m_{0^{'}}^2-m_{0^{''}}^2}\equiv \frac{18}{\triangle  
}\simeq 27\;
{\rm GeV}^{-2},
\eeq
where the difference $\triangle \equiv m_{0^{'}}^2-m_{0^{''}}^2$ is  
determined
by a distance between the parallel Regge trajectories for the  
$\omega $ and
$\phi $ resonances, which are described by the straight lines  
$J=0.59+0.84M^2$
and $J=0.04+0.84M^2,$ respectively, so that $\triangle \simeq 0.65$  
GeV$^2.$
We note further that the meson nonets may be arranged in the pairs  
of nonets
which have equal parity but different spins (which differ by 2),
e.g.,\footnote{For the scalar meson nonet, we use the $q\bar{q}$  
assignment
suggested by the authors in ref. \cite{linear,enigmas}.}

1 $^3P_0$ $J^{PC}=0^{++},$ $a_0(1320),$ $f_0(1300),$ $f_0(1525),$  
$K_0^\ast
(1430),$

1 $^3P_2$ $J^{PC}=2^{++},$ $a_2(1320),$ $f_2(1270),$ $f_2^{'}(1525),$ 
$K_2^\ast (1430),$ \\

1 $^3D_1$ $J^{PC}=1^{--},$ $\rho (1700),$ $\omega (1600),$ $K^\ast  
(1680),$
(no $\phi $ candidate),

1 $^3D_3$ $J^{PC}=3^{--},$ $\rho _3(1700),$ $\omega _3(1600),$  
$\phi _3(1850),$
$K_3^\ast (1780),$ \\

3 $^1S_0$ $J^{PC}=0^{-+},$ $\pi (1770),$ $\eta (1760),$ $K(1830),$  
(no $\eta ^
{'}$ candidate),

1 $^1D_2$ $J^{PC}=\!2^{-+},$ $\pi _2(1670),$ $K_2(1770),$ (no $\eta  
,\eta {'}$
candidates), \\
and occupy the mass interval of an individual nonet but have 18  
isospin degrees
of freedom in this interval, i.e., twice as much as that for an  
individual
nonet. Moreover, as the temperature gets closer to the critical one  
of chiral
symmetry restoration, we expect the chiral partners (the states  
with equal
isospin but different parity) have equal masses and form parity  
doublets. The
work of DeTar and Kogut \cite{DTK} shows convincingly that the  
``screening
masses'' of chiral partners are different below and become equal  
above a common
$T_c.$ This work was carried out for four quark flavors. Similar  
results were
obtained for two flavors by Gottlieb {\it et al.} \cite{Got}. In these 
calculations, the chiral partners were $(\pi ,\sigma ),$ $(\rho  
,a_1)$ and
$(N(\frac{1}{2}+),N(\frac{1}{2}-)).$ Thus, we expect the correct  
density of
states per unit mass interval to be twice as much as that for an  
individual
nonet, and hence, the correct normalization constant is
\beq
C\simeq 54\;{\rm GeV}^{-2}.
\eeq
Once the mass spectrum of a nonet (with a given fixed spin) is  
established to
be linear, one may take into account different nonets with  
different spins in
Eqs. (3),(4). As shown in ref. \cite{spectrum}, since the particle  
spin is
related to its mass, $J_i\sim \alpha ^{'}m_i^2,$ $\alpha ^{'}$ being a 
universal Regge slope, the spin degeneracy turns out to be  
proportional to the
mass squared, and the account for different nonets results in the  
following
mass spectrum,
\beq
\tau ^{'}(m)=C^{'}m^3,\;\;\;C^{'}=2\alpha ^{'}C\simeq 90\;{\rm GeV}^{-4},
\eeq
which is the actual resonance spectrum of hadronic matter and  
should lead,
through (11), to the equation of state
\beq
p,\rho \sim T^8,\;\;\;p=\rho /7.
\eeq
Bebie {\it et al.} \cite{Bebie} have calculated the ratio $\rho /p$  
directly
from Eqs. (3),(4), with all known hadron resonances with the masses  
up to 2 GeV
taken into account, and found that the curve $\rho /p$ first  
decreases very
quickly and then saturates at the value of $\rho /p\simeq 7,$ as  
read off from
Fig. 1 of ref. \cite{Bebie}, in agreement with (21).

In order to show that the obtained normalization constant is  
correct, we note
that the number of states with the masses up to $M,$ given by the  
mass spectrum
(20), is
\beq
N(M)=\frac{C^{'}M^4}{4}\simeq 22.5\;(M,\;{\rm GeV})^4.
\eeq
For, e.g., $M=1.25$ Eq. (22) gives
\beq
N(1.25)\simeq 55.
\eeq
The masses up to 1.25 GeV have the members of the pseudoscalar and  
vector meson
nonets, and the $h_1(1170),$ $b_1(1235)$ and $a_1(1260)$ mesons,  
the mass of
the latter was indicated by the recent Particle Data Group as 1.23 GeV 
\cite{data1}. We do not include the scalar mesons $a_0(980)$ and  
$f_0(980)$
which seem to be non-$q\bar{q}$ objects \cite{enigmas}, but may  
include the $f_
0(1300)$ meson which has the mass lying in the interval $1-1.5$  
GeV, according
to the recent Particle Data Group. Thus, we have 9+27+1+9+9+1=56  
actual mesonic
species having the masses up to 1.25 GeV, in excellent agreement  
with (23).

For $M=1.7$ GeV, Eq. (22) gives
\beq
N(1.7)\simeq 188.
\eeq
As seen in the Meson Summary Table \cite{data}, the masses up to  
1.7 GeV have
the members of the following nonets: 1 $^1S_0,$ 1 $^3S_1,$ 1 $^1P_1,$ 1 
$^3P_0,$ 1 $^3P_1,$ 1 $^3P_2,$ 2 $^1S_0,$ 2 $^3S_1.$ Therefore, one has 
\beq
(20\;{\rm spin\;states})\times (9\;{\rm isospin\;states})=180\;{\rm  
states},
\eeq
in good agreement with the result (24) given by a cubic spectrum.

For $M=2$ GeV, Eq. (22) gives
\beq
N(2)\simeq 360.
\eeq
The masses up to 2 GeV have the members of all the nonets indicated in 
\cite{data} except for the 1 $^3F_4$ and 2 $^3P_2$ nonets. In this  
case, one
has
\beq
(41\;{\rm spin\;states})\times (9\;{\rm isospin\;states})=369\;{\rm  
states},
\eeq
again in good agreement with the result (26) given by a cubic  
spectrum. Thus,
we consider the cubic spectrum (20) as granted by the actual  
experimental meson
spectrum.

Now, as the actual meson resonance spectrum is established, one uses this
spectrum in Eqs. (5),(6), in order to obtain the equation of state  
of hot
hadronic (mesonic) matter. First we note that one may approximate  
the particle
statistics by the Maxwell-Boltzmann one, because of the richness of the 
spectrum $\tau (m)\sim m^3$ at large $m;$ then the sum over $r$ in  
Eqs. (5),(6)
may be approximated by $\pi ^4/90=\zeta (4)\equiv \sum _r1/r^4,$  
which is the
asymptotic form of this sum for $T>>m.$ One obtains, therefore,
\beq
p=\frac{\pi ^2}{30}T^4+\frac{C^{'}\pi ^2}{180}T^4\int  
dm\;m^3\left(\frac{m}{T}
\right)^2K_2\left(\frac{m}{T}\right),
\eeq
$$\rho =T\frac{dp}{dT}-p,$$ where we have separated out the  
contribution of
the pions which may well be treated as massless at temperatures  
$\sim 150$ MeV,
and taken into account the remaining particle species by an  
integration over
$m$ with the resonance spectrum (20), which therefore has the lower  
limit
$\sim 0.5\;{\rm MeV}\simeq m_K.$ We note that this treatment of the  
hadronic
resonance gas corresponds, in view of (3),(5), to a collection of free 
(non-interacting) particles, which is completely justified at lower 
temperatures, since chiral symmetry suppresses the interactions of  
low energy
Goldstone bosons both among themselves and with massive hadrons,  
but becomes
an approximation as the temperature increases. Since the main  
contribution to
integrals of the type (28) is given by the mass region in which  
$m>>T,$ one
may extend the upper limit of integration to infinity and neglect  
the lower
limit, and obtain, through the formula \cite{GR}
$$\int _0^\infty dx\;x^\mu K_\nu (ax)=2^{\mu -1}a^{-\mu -1}\Gamma  
\left(\frac{
1+\mu +\nu }{2}\right)\Gamma \left(\frac{1+\mu -\nu }{2}\right),$$
\beq
p\simeq \frac{\pi ^2}{30}T^4\left[\;1+\left(\frac{T}{160\;{\rm  
MeV}}\right)^4
\;\right],
\eeq
\beq
\rho \simeq \frac{\pi ^2}{10}T^4\left[\;1+\frac{7}{3}\left(\frac{T}{160\;
{\rm MeV}}\right)^4\;\right].
\eeq
The formulas (29),(30) represent the equation of state of hot  
mesonic matter.

If one restricts himself to the lower mass resonances (i.e., the  
lower spin,
$J=0,1,$ nonets) alone, one may use the linear spectrum (14) with  
$C$ defined
in (19) and obtain
\beq
p=\frac{C\pi ^2}{180}T^4\int dm\;m\left(\frac{m}{T}\right)^2
K_2\left(\frac{m}{T}\right)\simeq \left(23\;{\rm GeV}^{-2}\right)T^6,
\eeq
in apparent agreement with the Shuryak's ``realistic'' equation of  
state (12).

The formulas (29),(30) have been obtained for the meson resonances  
alone. There
is no difficulty of principle to consider the baryon resonances in  
a similar
way, with the inclusion of two chemical potentials, for both  
conserved net
baryon number and strangeness. As we have checked in ref.  
\cite{spectrum}, a
mass spectrum of the $SU(3)$ baryon multiplets is linear, as well  
as for the
meson nonets, although to establish its correspondence to the  
Gell-Mann--Okubo
formulas is more difficult than for a meson nonet, since these  
formulas are
linear in mass for baryons (more detailed discussion is given in ref. 
\cite{spectrum}). Recent result of Kutasov and Seiberg \cite{KS}  
shows that
the numbers of bosonic and fermionic states in a non-supersymmetric 
tachyon-free string theory must approach each other as increasingly  
massive
states are included. The experimental hadronic mass spectrum shows  
that in the
mass range $\sim 1.2-1.7$ GeV, the number of baryon states nearly  
keeps pace
with that of meson states \cite{FR} (and, therefore, is well  
described by the
same cubic spectrum as for the mesons, Eq. (20)). Above $\sim 1.7$  
GeV, the
number of the observed baryons begins to outstrip that of the  
mesons, and then
greatly surpasses the latter at higher energies (indicating,  
therefore, that
the baryon resonance spectrum grows faster than (20) in this mass  
region, since
the cubic spectrum (20) describes the meson resonances well, up to,  
at least,
2 GeV, as we have seen in Eqs. (23)-(27)). The explanation of this  
behavior of
the experimental resonance spectrum was found by Cudell and Dienes  
in a naive
hadron-scale string picture \cite{CD}: the ratio of the numbers of  
the baryon
and meson states should, in fact, oscillate around unity, with the mesons
favored first, then baryons, then mesons again, etc. Keeping in mind this
picture, we may assume that the ``in-average'' baryon resonance  
spectrum has
the same form, Eq. (20), as the meson resonance one. If one now  
neglects, for
simplicity, the baryon number and strangeness chemical potentials (i.e., 
considers the case of both zero net baryon number and strangeness),  
and takes
into account the baryon resonances along with the meson ones by the mass 
spectrum (20), one will obtain the same formulas (28) but with an  
extra term in
the r.h.s. which differs from the second term of Eq. (28) by the  
factor 7/8
(since for the baryons, the sum over $r$ in Eqs. (5),(6) is  
approximated by
$7\pi ^2/720=7/8\;\zeta (4)=\sum _r(-1)^{r+1}/r^4).$ These formulas will
further reduce to the relations
\beq
p\simeq \frac{\pi ^2}{30}T^4\left[\;1+\left(\frac{T}{140\;{\rm  
MeV}}\right)^4
\;\right],
\eeq
\beq
\rho \simeq \frac{\pi ^2}{10}T^4\left[\;1+\frac{7}{3}\left(\frac{T}{140\;
{\rm MeV}}\right)^4\;\right],
\eeq
which represent the equation of state of hot hadronic (both mesonic and 
baryonic) matter.

Comparison of Eqs. (29),(32) with (1) shows that the agreement  
between the both
approaches, (i) the low-temperature expansion for the pion  
interacting gas, and
(ii) the hadronic resonance gas, is very good, if the latter is  
treated with
the help of the realistic mass spectrum, Eq. (20).

We shall use Eq. (1) and the corresponding formula for the energy  
density,
\beq
\rho \simeq \frac{\pi ^2}{10}T^4\left[\;1+\frac{7}{3}\left(\frac{T}{150\;
{\rm MeV}}\right)^4\;\right],
\eeq
as the equation of state of hot hadronic matter, in agreement with  
the results
(29),(32) provided by the hadronic resonance spectrum, Eq. (20). It  
is seen in
(34) that the ratio $\rho /T^4$ rapidly grows with $T,$ proportionally to
$T^4.$ This behavior is to be expected, since the value $\rho  
/T^4=\pi ^2/10
\simeq 1$ characteristic of a gas of the pions is small compared to  
the value
$\rho /T^4=47.5\pi ^2/30\approx 15.6$ which pertains to the plasma  
of free
quarks and gluons (47.5 being the number of degrees of freedom in  
the plasma
with $N_c=N_f=3).$ The energy density stored in the excited states  
reaches the
energy density of the pionic component at $T\simeq 120$ MeV, as  
follows from
(34), in agreement with the value $\simeq 125$ MeV obtained by  
Bebie {\it et
al.} \cite{Bebie} through direct calculation with the help of Eq.  
(4). The
quark-gluon plasma value $\rho \simeq 15.6\;T^4$ is reached at  
$T\simeq 235$
MeV. However, the energy densities of the hadron and quark-gluon  
phases need
not match at the critical point, since the phase transition may liberate 
latent heat. The only pressure in both phases should match at the  
critical
temperature, for which, therefore, a naive estimate may be obtained  
by equating
(1) with the plasma's $p=47.5\pi ^2/90\;T^4:$ $T_c\simeq 290$ MeV.  
The phase
transition from the hadron to the quark-gluon phase is expected to  
occur at
much lower temperature. Indeed, the approximation of the hadronic  
resonance gas
by a collection of free particles is justified up to temperatures  
$\sim 150$
MeV. Above this point, the interaction among the constituents of the gas 
rapidly grows with $T$ and inelastic collisions become increasingly  
important.
Moreover, direct calculation, with the help of Eqs. (3),(4), made  
by Bebie
{\it et al.} \cite{Bebie} shows that the mean distance between two  
particles
of the gas reaches the value $d=1$ fm at $T\simeq 190$ MeV,  
indicating that
Eqs. (3),(4) pack the hadrons very densely there and suggesting,  
therefore,
that color cannot remain confined much beyond this point. Thus, the  
equation
of state (1),(34) may be an adequate representation of (3),(4) up to 
temperatures $\sim 190$ MeV. The phase transition to the  
quark-gluon plasma is
expected to occur at the same temperatures, $\sim 180-190$  
MeV.\footnote{The
value of the critical temperature of the chiral symmetry restoration 
transition, to be associated with the hadron to quark-gluon one,  
calculated
with the contribution of the massive states being taken into account, is 
\cite{chiral} $\sim 190$ MeV if only the meson resonances, and  
$\sim 175$ MeV
if both the baryon and meson resonances are considered, in  
agreement with the
results obtained previously by Gerber and Leutwyler \cite{GL}.} We  
conclude,
therefore, that the equation of state (1),(34) is the realistic  
equation of
state of hot hadronic matter in the whole temperature range of the  
latter: $0-
T_c,$ with $T_c$ being the critical temperature of the hadron to  
quark-gluon
phase transition. In the following section we shall build a model for the
thermodynamics of this transition.

\section{Phase transition to the quark-gluon plasma}
In the simplest schematic model the quark-gluon plasma is described  
by the
equation of state
\beq
p=\frac{47.5\pi ^2}{90}T^4-B,\;\;\;\rho =\frac{47.5\pi^2}{30}T^4+B,
\eeq
where $B$ is the bag constant, and 47.5 the effective degeneracy of  
8 gluons
and 3 massless quarks. In going from the hadron phase (and its  
nonperturbative
vacuum) to the deconfined quark-gluon phase (with perturbative vacuum) we
restore conformal invariance. Consequently \cite{Brown}, we should  
use the
value of $B$ given by the conformal anomaly:
\beq
B=\frac{1}{4}\langle 0|\theta _\mu ^\mu |0\rangle =-\frac{1}{4}\langle 0|
\frac{\beta (g)}{2g}(G_{\mu \nu}^a)^2|0\rangle .
\eeq
Here $\theta _\mu ^\mu $ is the trace of the energy-momentum tensor and
$\langle 0|(G_{\mu \nu }^a)^2|0\rangle $ is the gluon condensate. The 
$\beta $-function of the renormalization group is given by \cite{renorm}
\beq
\beta (g)=-\frac{g^3}{16\pi ^2}\left(11-\frac{2}{3}N_f\right)
\eeq
for $N_f$ flavors. The value of the gluon condensate,
\beq
\frac{g^2}{4\pi ^2}\langle 0|(G_{\mu \nu }^a)^2|0\rangle \simeq  
\left(330\;
{\rm MeV}\right)^4,
\eeq
is well determined by the states of charmonium using QCD sum rules
\cite{sumrules}. For three flavors, we then find
\beq
B^{1/4}\simeq 240\;{\rm MeV},
\eeq
and Eqs. (35) reduce to
\beq
p\simeq \frac{47.5\pi ^2}{90}T^4\left[\;1-\left(\frac{160\;{\rm  
MeV}}{T}\right)
^4\;\right],\;\;\;
\rho \simeq \frac{47.5\pi ^2}{30}T^4\left[\;1+\frac{1}{3}\left(
\frac{160\;{\rm MeV}}{T}\right)^4\;\right].
\eeq
If the hadron phase is described, e.g., by Eqs. (29),(30), a simple  
equating of
the pressure in both phases (Eqs. (29) and (40)) gives two real  
positive roots:
\beq
T_1\approx 166\;{\rm MeV,}\;\;\;T_2\approx 308\;{\rm MeV.}
\eeq
In the temperature range $T_1\leq T\leq T_2,$ the curve $p=p(T)$ for the 
quark-gluon phase goes above the corresponding curve for the hadron  
phase, but
at $T>T_2,$ the situation changes: the quark-gluon curve goes below  
the hadron
one. In this picture, therefore, the hadron phase is again  
thermodynamically
favored at high temperatures, and the existence of the quark-gluon  
phase is
restricted to a limited range of temperatures, which is clearly  
unphysical. The
reason for such unphysical behavior predicted by Eqs. (29),(40) is,  
except for
the applicability of Eqs. (29),(30) limited by $T\sim 190$ MeV, the  
effective
degeneracy in the hadron phase which grows, according to (29), as  
$g_{eff}\sim
T^4.$ As shown by Brown {\it et al.} \cite{Brown}, the free energy of an 
interacting system of massless microscopic constituents (and hence,  
of the
effective degrees of freedom which are intended to mimic these  
interactions)
should be greater than the free energy of the noninteracting microscopic 
constituents at any temperature. When these interactions are  
realized in terms
of massless hadrons, this constraint implies that the number of the  
effective
(hadronic) degrees of freedom should always be less than 47.5 and that it
should {\it approach} 47.5 in the high-temperature limit. In other  
words, the
number of degrees of freedom in the effective excitations (hadrons)  
should not
exceed that of the underlying quarks and gluons. Specifically, the  
hadronic
resonances are intended to simulate attractive channels of the  
microscopic
constituents. The stability of the trivial microscopic vacuum at zero
temperature against the production of infinitely many particles  
requires the
existence of repulsive channels for the microscopic constituents,  
which is not
taken into account in Eqs. (1),(29),(32). At the one-body level,  
such repulsion
is most easily simulated by associating an excluded volume with  
each hadron
\cite{Brown}. However, the most naive assumption of a temperature  
independent
excluded volume represents an ``overkill'' in the high-temperature  
description
of the now massless hadrons, since it leads to an effective  
degeneracy which
vanishes as the temperature grows up \cite{Brown}. A more realistic  
description
of the hadron phase can be obtained using an excluded volume which  
decreases
like $1/T^3$ for large $T,$ which enforces the limitation of the  
number of the
effective degrees of freedom by a fixed number, which may be made  
taken a
desired value (e.g., 47.5) by the corresponding choice of the  
parametrization
for a temperature dependent excluded volume \cite{Brown}. An alternative
approach is to restrict ourselves to a finite number of hadronic  
states, which
should be chosen to be 47.5, since, according to the arguments  
given above, we
should have 47.5 effective massless degrees of freedom in the  
hadron phase at
high temperature. For our present semiquantitative purposes, this  
approach
provides an adequate alternative to the use of excluded volumes.

The most natural way to restrict the effective degeneracy in the  
hadron phase
(to 47.5) is truncating the resonance spectrum (20) when the number  
of the
effective hadronic degrees of freedom is 47.5. This makes certain
sense from the physical point of view, similar to the vibrations of  
a Debye
solid where the phonon spectrum is truncated when the number of  
phonons is
$3N,$ with $N$ being the number of atoms. It is seen in (22) that  
$M\simeq 1.2$
GeV corresponds to $N=47.5;$ we consider, therefore, the following  
expression
for the pressure in the hadron phase with 47.5 effective degrees of  
freedom:
\beq
p=\frac{\pi ^2}{30}T^4+\frac{C^{'}\pi ^2}{180}T^4\int  
_{0.5}^{1.2}dm\;m^3\left(
\frac{m}{T}\right)^2K_2\left(\frac{m}{T}\right),
\eeq
where we have, as previously, separated out the contribution of  
(massless)
pions. It is now seen that at $T\stackrel{>}{\sim }1.2$ GeV, this  
formula
reduces, through $K_2(x)\sim 2/x^2,\;x<<1,$ to
\beq
p=\frac{\pi ^2}{30}T^4+\frac{C^{'}\pi ^2}{90}T^4\int _{0.5}^{1.2}dm\;m^3;
\eeq
since $C^{'}\int _{0.5}^{1.2}dm\;m^3=44.5\;(=47.5-3$ of the pions),  
it follows
from (43) that at high enough temperature,
\beq
p=\frac{47.5\pi ^2}{90}T^4,
\eeq
i.e., we have 47.5 effective massless degrees of freedom in the  
hadron phase.
Further comparison of Eqs. (40) and (44) shows that the transition to the
deconfined quark-gluon phase is still prohibited by the bag  
constant $B:$ the
pressure curve (40) goes below the corresponding (44) at high $T$ ,  
and hence
the hadron phase is thermodynamically favored there, even if the  
curves (40)
and (44) intersect at some lower temperature. Thus, one sees that  
even in the
model with the restricted effective degeneracy in the hadron phase, the 
deconfinement transition is {\it absent.} This conclusion is in  
agreement with
lattice gauge calculations which show the only phase transition,  
viz., the
restoration of chiral symmetry \cite{DTK,Got}, not deconfinement.  
As stressed
out in ref. \cite{Brown}, the deconfinement transition for pure  
glue, used in
lattice calculations as a model problem for many years, is a fictitious 
transition; it is completely changed by the introduction of light  
mass quarks.
Moreover, the results of lattice gauge calculations indicate that  
the gluon
condensate is essentially unchanged during the chiral restoration  
transition
\cite{CDG}. Even at a temperature as high as $\sim 290$ MeV, lattice 
calculations find about half as much gluon condensate as at zero  
temperature
\cite{Lee}. Apparently, the conformal anomaly (and the bag constant  
associated
with the former) is relatively unaffected by what happens to the quark
condensate. The same conclusion was reached by Brown \cite{Brown1} in the
connection with the variations of a chemical potential rather than  
temperature.
Thus, even at temperatures well above the critical one of chiral symmetry
restoration, $T_\chi \;(\sim 190$ MeV), gluons remain condensed.  
Consequently,
although the mesons can be regarded as quark-antiquark pairs, each  
quark must
be connected with an antiquark by a ``string'' (i.e., a line  
integral of the
gauge vector potential), in order to preserve gauge invariance. It  
is difficult
to include the important consequences of such quark-antiquark  
correlations in
the thermodynamics of the gluon condensed system, unless one keeps  
regarding
the mesons as the correct effective degrees of freedom even above  
$T_\chi .$
Since our equation of state (42) does regard the mesons as the effective 
degrees of freedom at any $T,$ we may model the chiral symmetry  
restoration
transition with the help of Eq. (42).

Campbell {\it et al.} \cite{CEO} have shown how to build the  
correct operator
scaling properties into a low-energy effective Lagrangian in the
Nambu-Goldstone (chirally-asym-  metric) sector. Brown and Rho \cite{BR} 
have noted that this implies (in the chiral limit in which the bare  
(current)
quark masses are zero) that
\beq
\frac{m_\rho ^\ast }{m_\rho }= \frac{m_N^\ast }{m_N}=\frac{f_\pi ^\ast }{
f_\pi }=\ldots =\Phi (T),
\eeq
where $\Phi (T)$ describes the common scaling of all hadronic  
properties. Here
the starred quantities denote in-medium values at finite  
temperature and/or
density. The order parameter for chiral symmetry breaking is the  
effective pion
decay constant, $f_\pi ^\ast $ \cite{CEO,BR}. One can associate  
$f_\pi ^\ast
\rightarrow 0$ with  with the melting of the quark condensate,  
$\langle \bar{
q}q\rangle ^\ast ,$ with increasing temperature. The form of $\Phi  
(T)$ was
justified in ref. \cite{BR}:
\beq
\Phi (T)=\left(\frac{\langle \bar{q}q\rangle ^\ast }{\langle  
\bar{q}q\rangle }
\right)^{1/3}.
\eeq
We shall, however, use a different form of $\Phi (T),$ viz.,
\beq
\Phi (T)=\frac{\langle \bar{q}q\rangle ^\ast }{\langle \bar{q}q\rangle },
\eeq
suggested by the Nambu model and shown by Koch and Brown \cite{KB}  
to model
lattice gauge data better than (46). With the temperature  
dependence of the
quark condensate found by Gerber and Leutwyler \cite{GL} and  
confirmed in our
recent paper \cite{chiral},
\beq
\frac{\langle \bar{q}q\rangle ^\ast }{\langle \bar{q}q\rangle }=\left(1-
\frac{T^2}{T_\chi ^2}\right),\;\;\;T_\chi \simeq 190\;{\rm MeV,}
\eeq
Eq. (45) leads to the relation
\beq
m^\ast =m\left(1-\frac{T^2}{T_\chi ^2}\right)
\eeq
for all hadronic species. With (49), Eq. (42) takes on the form
\beq
p=\frac{\pi ^2}{30}T^4+\frac{C^{'}\pi ^2}{180}T^4\int  
_{0.5}^{1.2}dm\;m^3\left(
\frac{m^\ast }{T}\right)^2K_2\left(\frac{m^\ast }{T}\right),
\eeq
which upon the integration reduces, through  
$C^{'}(1.2^4-0.5^4)/4=44.5,$ to
$$p=\frac{\pi ^2}{90}T^4\left[\;\frac{3}{2}\left(\frac{m_\pi  
}{T}\right)^2K_2
\left(\frac{m_\pi  
}{T}\right)+89\;\frac{y^5K_5(y)-x^5K_5(x)-6y^4K_4(y)+6x^4
K_4(x)}{x^4-y^4}\;\right],$$
\beq
x\equiv \frac{1.2\Phi (T)}{T},\;\;\;y\equiv \frac{0.5\Phi (T)}{T},
\eeq
where $\Phi (T)$ is given in (47), and we have recovered the pion  
contribution
term with the finite pion mass. Temperature dependence of the effective
degeneracy, $g_{eff}\equiv p/p_{SB},$ with $p_{SB}\equiv 47.5\pi  
^2/90\;T^4,$
as well as those of $\rho /\rho _{SB},$ with $\rho _{SB}\equiv  
3p_{SB},$ and
$p/\rho $ $(\simeq c_s^2),$ are shown in Figs. 1-3, respectively.  
One sees that
in the model considered here, chiral symmetry restoration  
corresponds to a
smooth crossover in the thermodynamic variables $p$ and $\rho .$  
Let us compare
these results with those of lattice gauge calculations. To date the most 
complete lattice gauge calculations on energy production for  
temperatures near
$T_\chi $ are those obtained by Kogut {\it et al.} \cite{KSW}. For  
calculations
involving two low mass quarks, they find rapid increases in both  
the quark and
gluon energy densities beginning at $\beta =5.30$ on a $6\times  
12^3$ lattice
$(\beta $ is related to the inverse coupling constant, $\beta  
=6/g^2).$ By
$\beta =5.35,$ the sum of the quark and gluon energy densities is  
essentially
blackbody (and corresponds to that of an ultrarelativistic  
Stefan-Boltzmann
gas), although in the case of gluons the accuracy of their  
calculation is not
good enough. They find ``a smooth $\rho _{u,d}/T^4$ curve with  
considerable
suppression [compared to blackbody] in the transition region.'' The  
measured
transition is a smooth crossover, in contrast to a sharper change  
obtained
previously with smaller lattices. Earlier calculations of the  
behavior of the
quark and gluon energy densities in this temperature region are  
summarized by
Petersson \cite{Pet} and, within their accuracy, are roughly  
consistent with
the results of Kogut {\it et al.} \cite{KSW}.

This smooth crossover is in agreement with the results obtained in  
this paper
for the model of the hadron to quark-gluon transition with the truncated 
resonance spectrum and the temperature dependent effective hadron  
masses. It is
seen in Figs. 1-3 that the crossover begins at $T\sim 140$ MeV  
(corresponding
to a dip in the $T$-dependence of $p/\rho $ at those temperatures  
seen in Fig.
3 \cite{Shu}), i.e., at $T\simeq 3/4\;T_\chi ,$ and finishes at  
$T\sim T_\chi
,$ in agreement with the predictions made by Brown {\it et al.}  
\cite{Brown}.
At $T>T_\chi \simeq 190$ MeV, the system is essentially blackbody.

Although the results obtained in this paper are semiquantitative,  
they agree
well with available lattice gauge data and reflect the main  
features of the
latter,\footnote{These features are seen, e.g., in Fig. 1 of the  
most recent
review on lattice gauge calculations \cite{Karsch}: the crossover  
starts at
$T\sim 150$ MeV and finishes at $T\sim 200$ MeV, the ratio $\rho  
/\rho _{SB}$
approaches unity from above as the temperature increases. With the  
account for
the error bars of this calculation, the system is essentially  
blackbody at $T$
just above 200 MeV.} i.e., a smooth crossover in the thermodynamic  
variables
taking place in a temperature range $\sim 50$ MeV, and the temperature 
dependent ratio $\rho /\rho _{SB}$ approaching unity from above as the 
temperature increases.

\section{Concluding remarks}
We have obtained the equation of state of hot hadronic matter, by  
taking into
account the contribution of the massive states with the help of the  
resonance
spectrum $\tau (m)\sim m^3$ justified in our previous papers. This
equation of state is in agreement with that provided by the  
low-temperature
expansion for the pion intracting gas. We have shown that in this  
picture the
deconfinement phase transition is absent, in agreement with lattice gauge
calculations which show the only phase transition of chiral symmetry 
restoration. The latter was modelled with the help of the  
restriction of the
number of the effective degrees of freedom in the hadron phase to that
of the microscopic degrees of freedom in the quark-gluon phase,  
through the
corresponding truncation of the hadronic resonance spectrum, and  
the decrease
of the effective hadron masses with temperature, predicted by Brown  
and Rho.
The results are in agreement with lattice gauge data, thereby  
confirming the
prediction of Brown and Rho, and show a smooth crossover in the  
thermodynamic
variables which begins at $T\simeq 3/4\;T_\chi $ and finishes at  
$T\simeq T_
\chi ,$ where $T_\chi \simeq 190$ MeV is the critical temperature  
for the
chiral symmetry restoration transition, i.e., takes place in a  
temperature
range $\sim 50$ MeV. At $T\cong T_\chi $ the system ends up in the  
chirally
symmetric phase and is essentially blackbody.

Although our results are in good agreement with lattice gauge data,  
we, along
with the authors of ref. \cite{Brown}, have at present no way of how to
decompose the hadron pressure (and energy density),  
$$p=\frac{g_{eff}\pi ^2}{
90}T^4,\;\;\;g_{eff}\rightarrow 47.5\;\;{\rm with}\;\;T\rightarrow  
\infty ,$$
into the separate quark and gluon components both above and below  
$T_\chi .$
Also, we are not clear about what happens above $T_\chi $ in a real  
world.
This is the matter of subsequent investigations.

\section*{Acknowledgements}
One of us (L.B.) wishes to thank E.V. Shuryak and G.E. Brown for  
very valuable
discussions on the hadronic resonance spectrum and the  
thermodynamics of hot
hadronic matter.

\newpage
\centerline{FIGURE CAPTIONS}
\bigskip
\bigskip
\bigskip
\bigskip
\hfil\break
Fig. 1. The ratio $p/p_{SB},$ with $p$ given in Eq. (51), as a  
function of
temperature.\hfil\break
\hfil\break
\hfil\break
\hfil\break
Fig. 2. The ratio $\rho /\rho _{SB}$ as a function of  
temperature.\hfil\break
\hfil\break
\hfil\break
\hfil\break
Fig. 3. The ratio $p/\rho $ as a function of temperature.

\bigskip
\bigskip
\begin{thebibliography}{9}
\bibitem{Shu} E.V. Shuryak, Phys. Rep. {\bf 61} (1980) 72
\bibitem{PGY} D.J. Gross, R.D. Pisarsky and L.G. Yaffe, Rev. Mod. Phys.
{\bf 53} (1981) 43
\bibitem{Mu} B. M\"{u}ller, {\it The Physics of the Quark-Gluon Plasma,}
(Springer-Verlag, Berlin, 1985)
\bibitem{CGS} J. Cleymans, R.V. Gavai and E. Suhonen, Phys. Rep.
{\bf 130} (1986) 217
\bibitem{QCD} E.V. Shuryak, {\it The QCD Vacuum,
Hadrons and the Superdense Matter,} (World Scientific, Singapore, 1988)
\bibitem{Lan} L.D. Landau, Izv. Akad. Nauk SSSR, seriya Fiz. {\bf 17}
(1953) 51
\bibitem{Mel1} G.A. Milekhin, Sov. Phys. JETP {\bf 8} (1959) 829
\bibitem{Shu1} E.V. Shuryak, Sov. J. Nucl. Phys. {\bf 16} (1973) 220
\bibitem{Hag} R. Hagedorn, Nuovo Cim. {\bf 35} (1965) 216;
Nuovo Cim. Suppl. {\bf 3} (1965) 147, {\bf 6} (1968) 311
\bibitem{Fra} S.C. Frautschi, Phys. Rev. D {\bf 3} (1971) 2821
\bibitem{FV} S. Fubini and G. Veneziano, Nuovo Cim. A {\bf 64} (1969) 811
\bibitem{FW} G.N. Fowler and R.M. Weiner, Phys. Lett. B {\bf 89}  
(1980) 394
\bibitem{ZS} O.V. Zhirov and E.V. Shuryak, Sov. J. Nucl. Phys. {\bf 21}
(1975) 443
\bibitem{Shu2} E.V. Shuryak, Phys. Rev. D {\bf 42} (1990) 1764
\bibitem{Shu3} E.V. Shuryak, Nucl Phys. A {\bf 522} (1991) 377c;
{\bf 533} (1991) 761
\bibitem{Shu4} E.V. Shuryak, Nucl Phys. A {\bf 536} (1992) 739
\bibitem{Wei} S. Weinberg, Phys. Rev. Lett. {\bf 17} (1966) 616
\bibitem{GL} P. Gerber and H. Leutwyler, Nucl. Phys. B {\bf 321} (1989)
387
\bibitem{BL} S.Z. Belenky and L.D. Landau, Sov. Phys. Uspekhi {\bf 62}
(1956) 1
\bibitem{BU} E. Beth and G.E. Uhlenbeck, Physica {\bf 3} (1936) 729
\bibitem{LL} L.D. Landau and E.M. Lifshitz, {\it Statistical Physics,}
Part 1, (Pergamon Press, Oxford, 1986), p. 236
\bibitem{Mel2} G.A. Milekhin, in {\it Proc. of International Conference
on Cosmic Rays,} (Nauka, Moscow, 1960)
\bibitem{SW} S. Sohlo and G. Wilk, Lett. Nuovo Cimento {\bf 13} (1975)
375
\bibitem{Gor} M.I. Gorenstein {\it et al.,} Phys. Lett. B {\bf 60} (1976)
283
\bibitem{And} B. Anderson {\it et al.,} Nucl. Phys. B {\bf 112} (1976)
413
\bibitem{spectrum} L. Burakovsky, L.P. Horwitz and W.C. Schieve,  
Hadronic
Resonance Spectrum: Power vs. Exponential Law. Experimental  
Evidence; to be
published
\bibitem{su4} L. Burakovsky and L.P. Horwitz, Gell-Mann--Okubo Mass  
Formula for
$SU(4)$ Meson Hexadecuplet, to be published
\bibitem{linear} L. Burakovsky and L.P. Horwitz, Mass Spectrum of a  
Meson
Nonet is Linear, to be published
%\bibitem{Bal} L.A.P. Bal\'{a}zs, Phys. Lett. B {\bf 120} (1983) 426
\bibitem{mancov}  L.P. Horwitz and C. Piron, Helv. Phys. Acta {\bf  
46} (1973)
316 \\ L.P. Horwitz, W.C. Schieve and C. Piron, Ann. Phys. (N.Y.)  
{\bf 137}
(1981) 306 \\ L.P. Horwitz, S. Shashoua and W.C. Schieve, Physica A  
 {\bf 161}
(1989) 300 \\ L. Burakovsky, Manifestly Covariant Relativistic  
Statistical
Mechanics as a Framework for Description of Realistic Physical  
Systems, Ph.D.
thesis (Tel-Aviv University, 1995), unpublished; L. Burakovsky,  
L.P. Horwitz
and W.C. Schieve, Mass - Proper Time Uncertainty Relation in a Manifestly
Covariant Relativistic Statistical Mechanics, to be published
\bibitem{cond} L. Burakovsky, L.P. Horwitz and W.C. Schieve, A New  
Relativistic
High Temperature Bose-Einstein Condensation; Phys. Rev. D, {\it in press}
%\bibitem{Stu} E.C.G. Stueckelberg, Helv. Phys. Acta {\bf 14} (1941)
%372, 588; {\bf 15} (1942) 23
%\bibitem{AHL} R. Arshansky, L.P. Horwitz and Y. Lavie, Found. Phys.
%{\bf 13} (1983) 1167
\bibitem{HW} H.A. Haber and H.E. Weldon, Phys. Rev. D {\bf 25} (1982) 502
%\bibitem{glim} L. Burakovsky and L.P. Horwitz, J. Phys. A {\bf 27}  
(1994) 4725
\bibitem{GMO} S. Okubo, Prog. Theor. Phys. {\bf 27} (1962) 949, {\bf 28} 
(1962) 24 \\ M. Gell-Mann and Y. Ne'eman, {\it The Eightfold Way,}  
(Benjamin,
N.Y., 1964)
\bibitem{enigmas} L. Burakovsky and L.P. Horwitz, Hadronic  
Resonance Spectrum
May Help in Resolution of Meson Nonet Enigmas, to be published
\bibitem{DTK} C. DeTar and J. Kogut, Phys. Rev. Lett. {\bf 59}  
(1987) 399;
Phys. Rev. D {\bf 36} (1987) 2828
\bibitem{Got} S. Gottlieb {\it et al.,} Phys. Rev. Lett. {\bf 59}  
(1987) 2247
\bibitem{Bebie} H. Bebie, P. Gerber, J.L. Goity and H. Leutwyler,  
Nucl. Phys.
B {\bf 378} (1992) 95
\bibitem{data1} Particle Data Group, Phys. Rev. D {\bf 50} (1994) 1429 
\bibitem{data} Particle Data Group, Phys. Rev. D {\bf 50} (1994) 1320 
\bibitem{GR} I.S. Gradshteyn and I.M. Ryzhik, {\it Table of  
Integrals, Series,
and Products,} (Academic Press, New York, 1980), p. 684, formula 16
\bibitem{KS} D. Kutasov and N. Seiberg, Nucl. Phys. B {\bf 358}  
(1991) 600
\bibitem{FR} P.G.O. Freund and J.L. Rosner, Phys. Rev. Lett. {\bf  
69} (1992)
765
\bibitem{CD} J.-R. Cudell and K.R. Dienes, Phys. Rev. Lett. {\bf  
69} (1992)
1324
\bibitem{chiral} L. Burakovsky and L.P. Horwitz, On the Thermodynamics of
Chiral Symmetry Restoration, to be published
\bibitem{Brown} G.E. Brown, A.D. Jackson, H.A. Bethe and P.M.  
Pizzochero,
Nucl. Phys. A {\bf 560} (1993) 1035
\bibitem{renorm} E.V. Shuryak, Phys. Lett. B {\bf 79} (1978) 135
\bibitem{sumrules} M.A. Shifman, A.I. Vainshtein and V.I. Zakharov,  
NUcl. Phys.
B {\bf 147} (1979) 385, 488 \\ L.J. Reinders, H. Rubinstein and S.  
Yazaki,
Phys. Rep. {\bf 127} (1985) 1
\bibitem{CDG} M. Campostrini and A. Di Giacomo, Phys. Lett. B {\bf  
197} (1987)
403
\bibitem{Lee} S.H. Lee, Phys. Rev. D {\bf 40} (1989) 2484
\bibitem{Brown1} G.E. Brown, Nucl. Phys. A {\bf 522} (1991) 397c
\bibitem{CEO} B.A. Campbell, J. Ellis and K.A. Olive, Phys. Lett. B  
{\bf 235}
(1990) 325; Nucl. Phys. B {\bf 345} (1990) 57
\bibitem{BR} G.E. Brown and M. Rho, Phys. Rev. Lett. {\bf 66} (1991) 2720
\bibitem{KB} V. Koch and G.E. Brown, Nucl. Phys. A {\bf 560} (1993) 345 
\bibitem{KSW} J.B. Kogut, D.K. Sinclair and K.C. Wang, Phys. Lett.  
B {\bf 263}
(1991) 101
\bibitem{Pet} B. Petersson, Nucl. Phys. A {\bf 525} (1991) 237c
\bibitem{Karsch} F. Karsch, Nucl. Phys. A {\bf 590} (1995) 367c
\end{thebibliography}
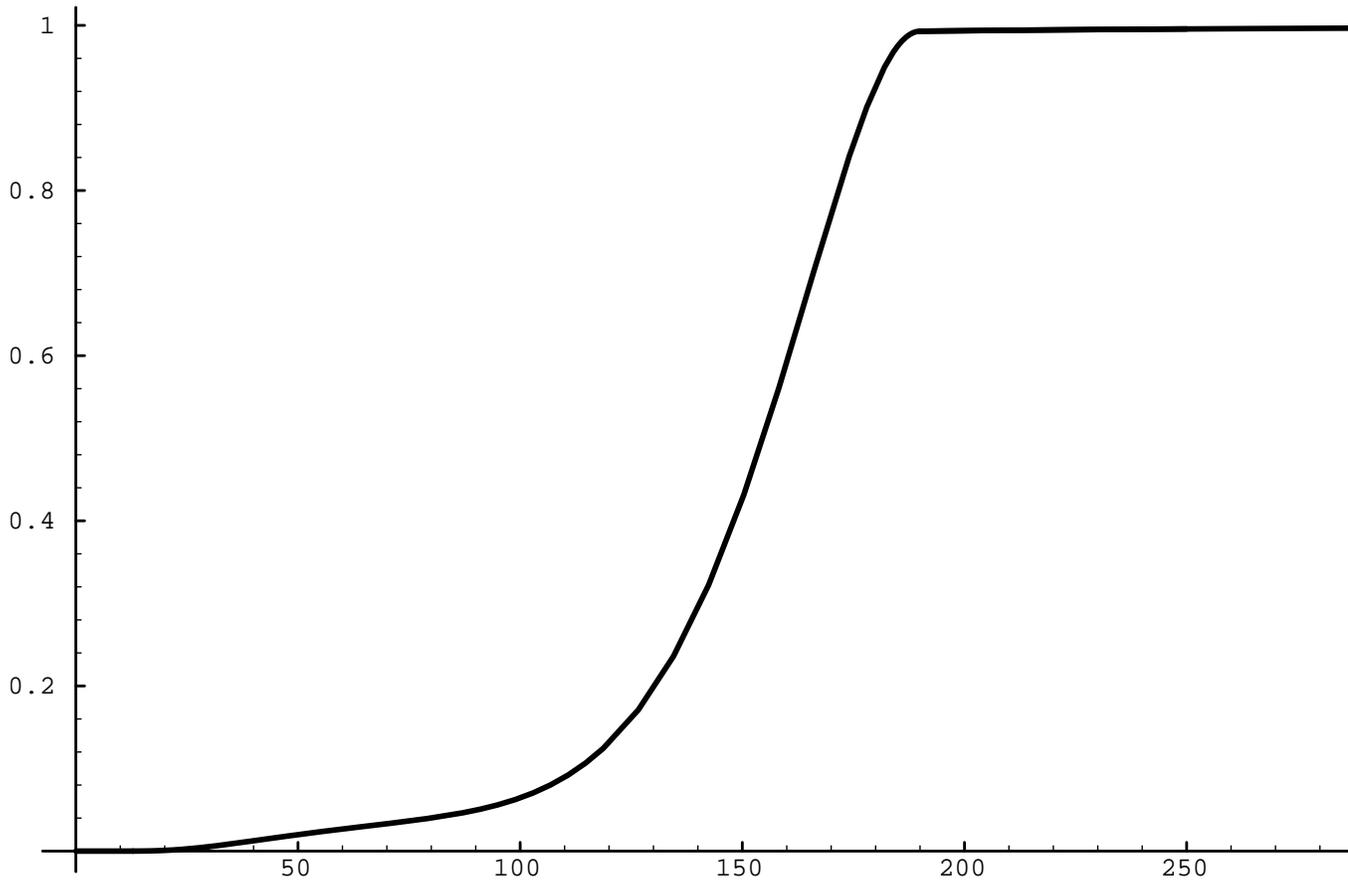
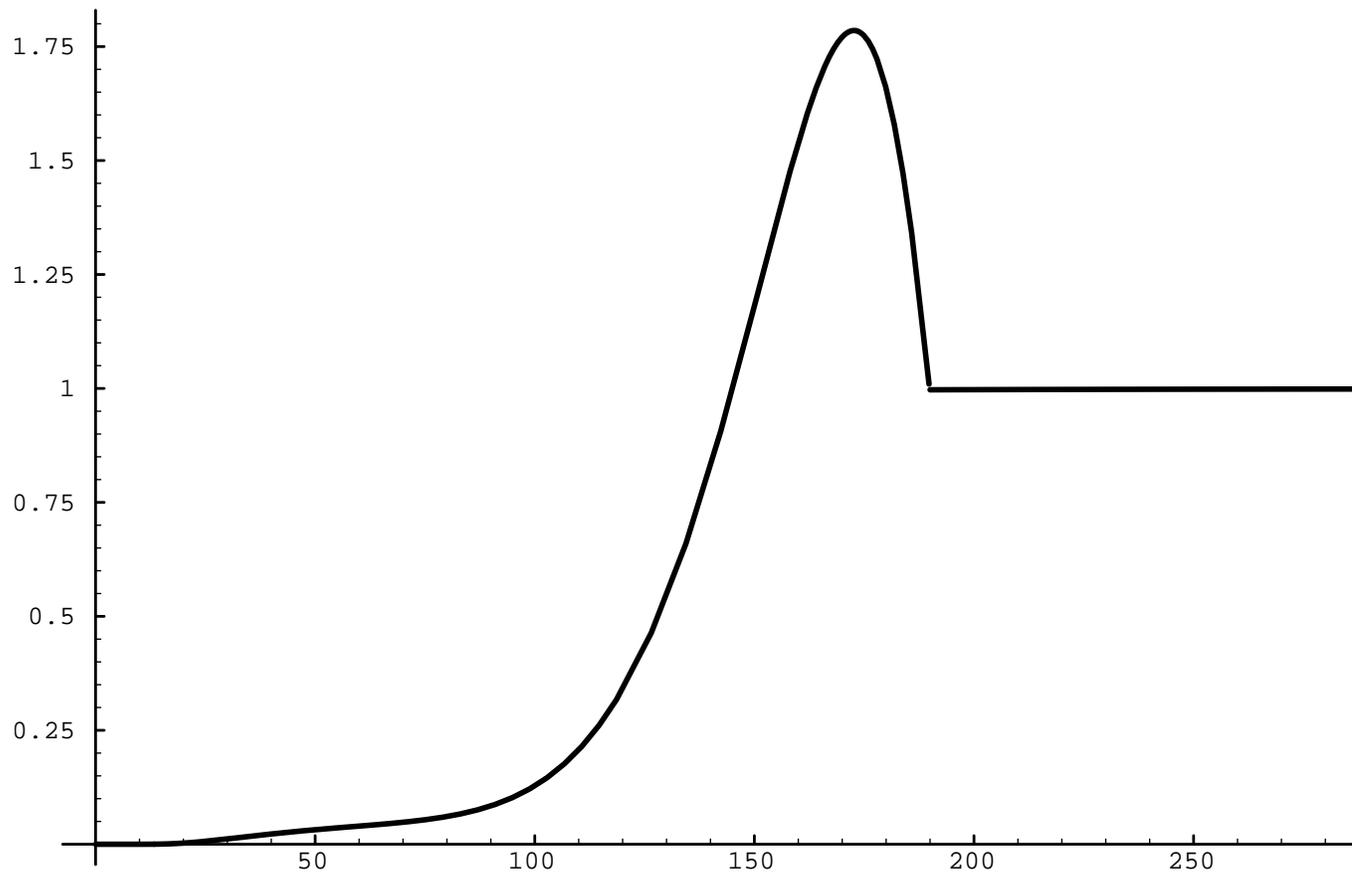
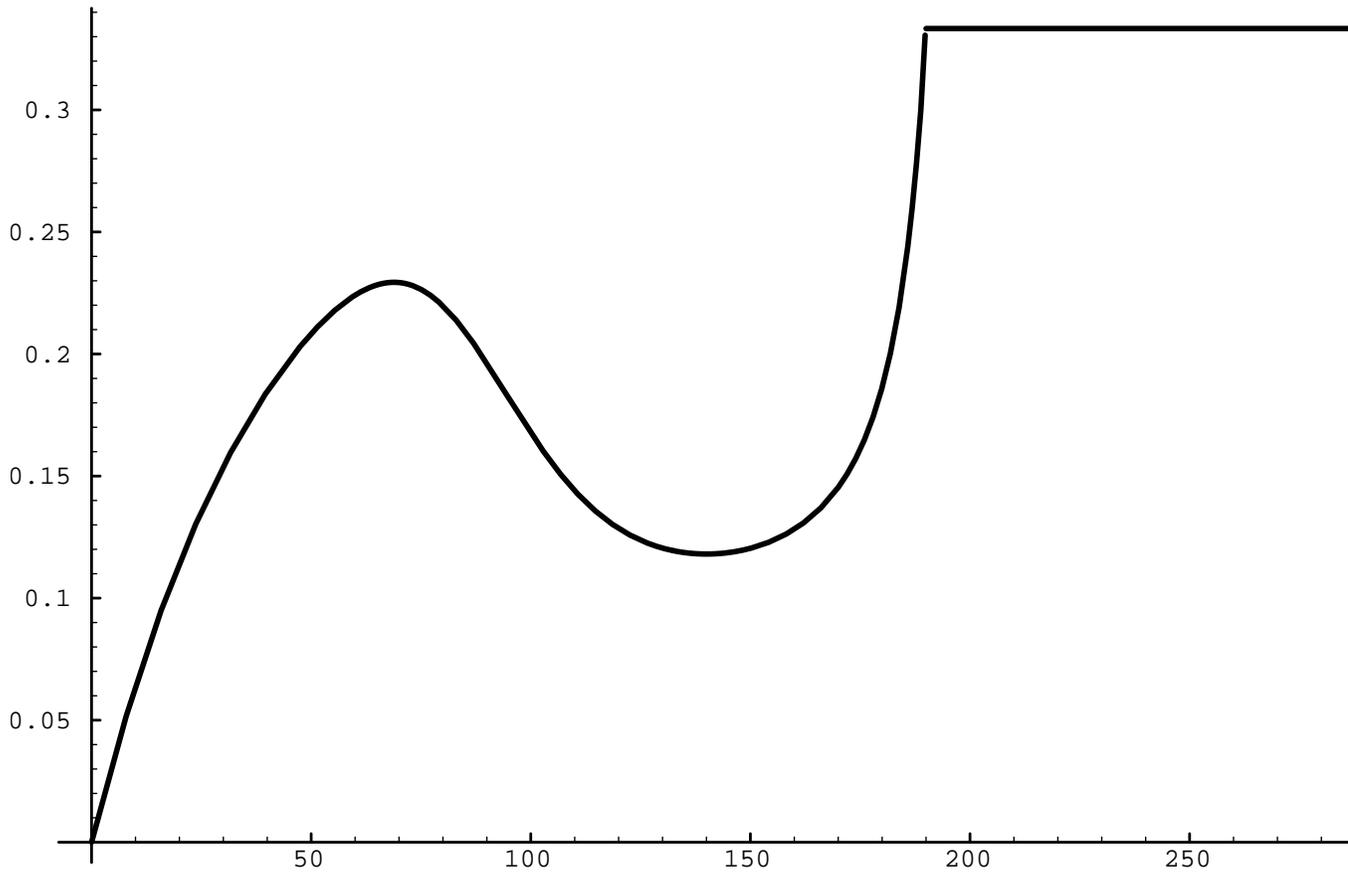
\end{document}